\documentclass{article}
\usepackage{spconf,amsmath,epsfig}

\let\OLDthebibliography\thebibliography
\renewcommand\thebibliography[1]{
  \OLDthebibliography{#1}
  \setlength{\parskip}{0pt}
  \setlength{\itemsep}{0pt plus 0.3ex}
}
\usepackage{cite}
\usepackage{amsmath,amssymb,amsfonts}
\usepackage{algorithmic}
\usepackage{graphicx}
\usepackage{textcomp}
\usepackage{xcolor}
\usepackage{subcaption}
\usepackage{multirow}
\usepackage{url}            % simple URL typesetting
\usepackage{booktabs}       % professional-quality tables
\usepackage{amsfonts}       % blackboard math symbols
\usepackage{microtype}      % microtypography
\def\BibTeX{{\rm B\kern-.05em{\sc i\kern-.025em b}\kern-.08em
    T\kern-.1667em\lower.7ex\hbox{E}\kern-.125emX}}

\begin{document}

\newcommand{\methodname}{{\tt{RLCP}}}
\title{RLCP: A Reinforcement Learning-based Copyright Protection Method for Text-to-Image Diffusion Model}

\name{
    % Authors
    Zhuan Shi\textsuperscript{\rm 1},
    Jing Yan\textsuperscript{\rm 1, \rm 2},
    Xiaoli Tang\textsuperscript{\rm 3},
    Lingjuan Lyu\textsuperscript{\rm 4},
    Boi Faltings\textsuperscript{\rm 1}
}
\address{
    % Affiliations
    \textsuperscript{\rm 1}EPFL, Switzerland\\
    \textsuperscript{\rm 2}ETH Zürich, Switzerland\\
    \textsuperscript{\rm 3}Nanyang Technological University, Singapore\\
    \textsuperscript{\rm 4}Sony AI\\
    \{zhuan.shi, jing.yan, boi.faltings\}@epfl.ch, xiaoli001@ntu.edu.sg, Lingjuan.Lv@sony.com
}

% \author{Zhuan Shi}
% \affiliatn{%
%   \institution{EPFL}
%   \city{Lausanne}
%   \country{Switzerland}
% }
% \email{zhuan.shi@epfl.ch}

% \author{Jin Yan}
% \affiliation{%
%   \institution{ETH Zürich}
%   \city{Zurich}
%   \country{Switzerland}
% }
% \email{christine.yifei.song@outlook.com}

% \author{Xiaoli Tang}
% \affiliation{%
%   \institution{Nanyang Technological University}
%   \city{}
%   \country{Singapore}
% }
% \email{xiaoli001@ntu.edu.sg}

% \author{Lingjuan Lyu}
% \affiliation{%
%   \institution{Sony AI}
%   \city{Zurich}
%   \country{Switzerland}
% }
% \email{Lingjuan.Lv@sony.com}

% \author{Boi Faltings}
% \affiliation{%
%   \institution{EPFL}
%   \city{Lausanne}
%   \country{Switzerland}
% }
% \email{boi.faltings@epfl.ch}

\maketitle

\begin{abstract}
The increasing sophistication of text-to-image generative models raises challenges in defining and enforcing copyright criteria. Existing methods like watermarking and dataset deduplication fall short due to the lack of standardized metrics and the complexity of addressing copyright issues in diffusion models. To tackle these challenges, we propose \methodname{}, a Reinforcement Learning-based Copyright Protection method for Text-to-Image Diffusion Models. Our approach introduces a novel copyright metric grounded in legal precedents and employs the Denoising Diffusion Policy Optimization (DDPO) framework to minimize copyright-infringing content while preserving image quality. A reward function based on our metric and KL divergence regularization ensures stable fine-tuning. Experiments on mixed datasets of copyright and non-copyright images show that \methodname{} effectively reduces copyright infringement risk without compromising output quality.

% The increasing sophistication of text-to-image generative models has led to complex challenges in defining and enforcing copyright infringement criteria and protection. 
% Existing methods, such as watermarking and dataset deduplication, fail to provide comprehensive solutions due to the lack of standardized metrics and the inherent complexity of addressing copyright infringement in diffusion models. 
% To deal with these challenges, we propose a \underline{R}einforcement \underline{L}earning-based \underline{C}opyright \underline{P}rotection(\methodname{}) method for Text-to-Image Diffusion Model, which minimizes the generation of copyright-infringing content while maintaining the quality of the model-generated dataset. 
% Our approach begins with the introduction of a novel copyright metric grounded in copyright law and court precedents on infringement. We then utilize the Denoising Diffusion Policy Optimization (DDPO) framework to guide the model through a multi-step decision-making process, optimizing it using a reward function that incorporates our proposed copyright metric. Additionally, we employ KL divergence as a regularization term to mitigate some failure modes and stabilize RL fine-tuning. 
% Experiments conducted on 2 mixed datasets of copyright and non-copyright images demonstrate that our approach significantly reduces copyright infringement risk while maintaining image quality.  
\end{abstract}

\begin{keywords}
Copyright Protection, Diffusion Model, Reinforcement Learning
\end{keywords}

\section{Introduction}
\label{sec:intro}

Recently, text-to-image diffusion models have garnered significant attention in research. These advanced methods \cite{balaji2023ediffi, rombach2022highresolution} have demonstrated exceptional capabilities in converting textual descriptions into highly accurate and visually coherent images. The advancements in these techniques have unlocked numerous possibilities for various downstream tasks, including image editing \cite{Avrahami_2022_CVPR, NEURIPS2020_4c5bcfec}, image denoising \cite{NEURIPS2020_4c5bcfec, xie2023diffusion}, and super-resolution \cite{ho2020denoising}. 

\begin{figure}[ht]
\centering
\includegraphics[width=0.9\columnwidth]{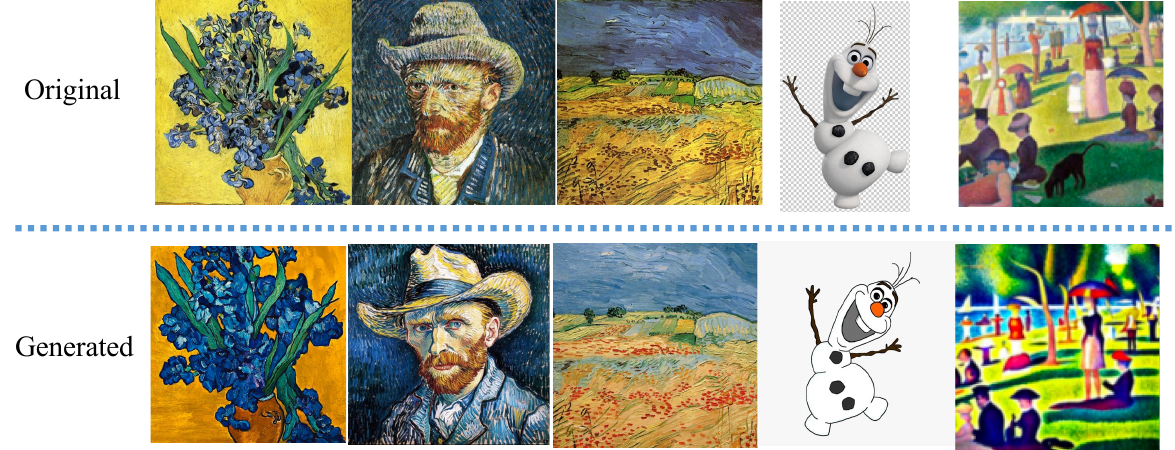}
\caption{Examples of data replication generated by text-to-image diffusion model.}
\label{fig:replication}
\end{figure}
While the progress in text-to-image generative models has profoundly impacted different industries, it also presents significant challenges for copyright protection. These models utilize extensive training data that may include copyrighted works, which they are sometimes capable of memorizing\cite{carlini2023extracting}. This ability can result in the production of images that closely resemble protected content in Figure \ref{fig:replication}, posing significant challenges to copyright protection\cite{elkin2023_copyright_privacy}. Recent legal cases, such as those involving Stable Diffusion\cite{rombach2022_ldm} and Midjourney\cite{mansour2023intelligent}, highlight concerns over the use of copyrighted data in AI training, where the models potentially infringe on the rights of numerous artists. These cases highlight a growing concern: Could the high-quality content synthesized by these generative AIs be excessively similar to copyrighted training data, potentially violating the rights of copyright holders?

Various methods have been proposed for source data copyright protection. One approach involves using unrecognizable examples\cite{gandikota2023erasing, zhang2023forgetmenot} that prevent models from learning key features of protected images either during inference or training stages. However, this method is highly dependent on the specific image and model, and it lacks general reliability.
Machine unlearning\cite{bourtoule2020machine, NEURIPS2019_cb79f8fa, nguyen2022survey} removes contributions of copyright data, aligning with the right to be forgotten, while dataset deduplication\cite{somepalli2022diffusion} helps reduce the risk of training sample memorization. 

Despite these efforts, existing copyright protection methods still have the following limitations: (1) They lack a standardized copyright metric that aligns with copyright laws and regulations, making it difficult to determine if generated images constitute copyright infringement; (2) These methods often prioritize performance on specific downstream tasks rather than focusing on general applicability, resulting in approaches that work well only on certain datasets but lack the versatility needed for broader use across diverse datasets.

To tackle these challenges, we propose a Reinforcement Learning-based Copyright Protection method (\methodname{}) for text-to-image diffusion models to reduce the possibility of generating copyright-infringing content. Specifically, inspired by Courts\footnote{\url{https://law.justia.com/cases/federal/appellate-courts/F2/562/1157/293262/}} in the US which employs two-part test to determine copyright violation, which contains an extrinsic test examining objective similarity in specific expressive elements, and an intrinsic test assessing subjective similarity from the perspective of a reasonable audience, we first propose copyright metric that mirror these legal standards by combining semantic and perceptual similarity. Then, we propose a novel framework that combines reinforcement learning with copyright infringement metrics. We leverage the Denoising Diffusion Policy Optimization framework to guide the model through a multi-step decision-making process, optimizing it using a reward function that incorporates our proposed copyright metric. Additionally, we employ KL divergence as a regularization term to mitigate some failure modes and stabilize RL fine-tuning.

Our main contributions are as follows:
\begin{enumerate}
    \item To the best of our knowledge, we are the first to propose a copyright metric that closely adheres to the procedures outlined in US copyright law, which could efficiently detect the substantial similarity between the original and generated images.
    \item We propose a novel framework that combines reinforcement learning with proposed copyright infringement metrics, which could reduce copyright infringement while preserving the quality of generated images.
    \item  Extensive experiments on 3 datasets highlight the significant superiority of the proposed \methodname{} in reducing the copyright loss while preserving quality of generated images than 4 baselines.
\end{enumerate}

\begin{figure*}[t!]
\centering
\includegraphics[width=0.85\linewidth]{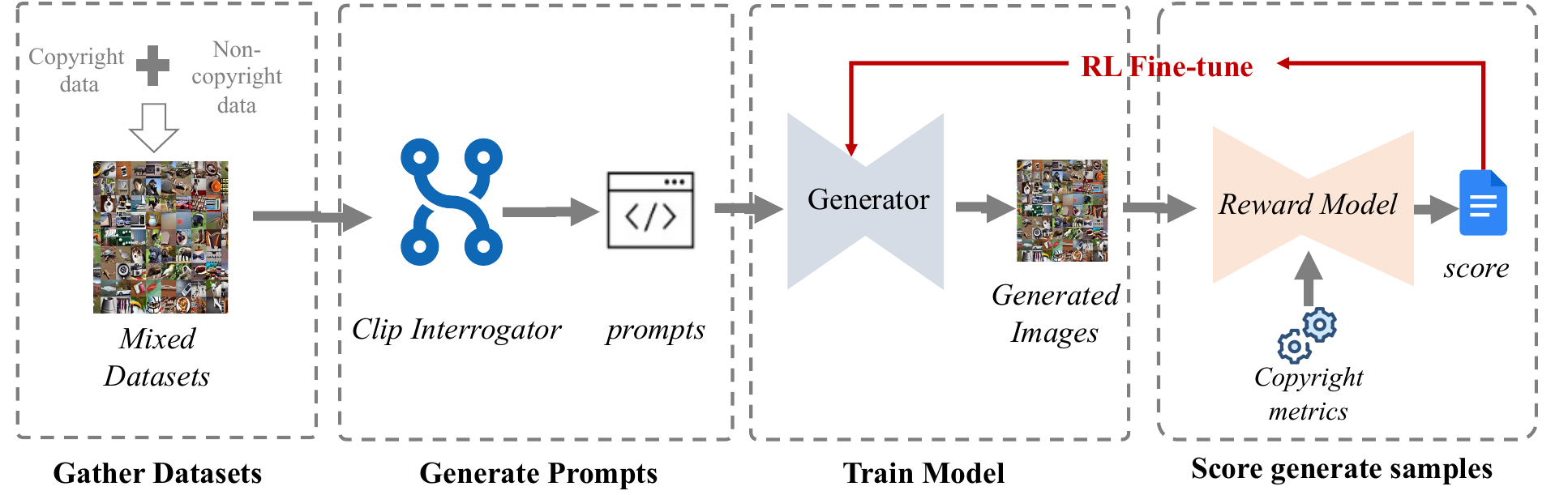}
\caption{Overview of \methodname{}.}
\label{fig:overview}
\end{figure*}

\section{Related Work}
\textbf{Text-to-Image Diffusion Models:} 
Recently, text-to-image diffusion models have garnered significant attention in research. These advanced methods \cite{balaji2023ediffi, rombach2022highresolution} have demonstrated exceptional capabilities in converting textual descriptions into visually coherent and realistic images with high accuracy. The advancements in these techniques have unlocked numerous possibilities for various downstream tasks, including image editing \cite{Avrahami_2022_CVPR, NEURIPS2020_4c5bcfec}, image denoising \cite{NEURIPS2020_4c5bcfec, xie2023diffusion}, and super-resolution \cite{ho2020denoising}. 

\textbf{Copyright Protection:} Several studies in the legal literature have examined copyright issues in machine learning and data mining, focusing primarily on potential infringements during the training phase.: (1) Concept Removal: To remove explicit artwork from large models, \cite{gandikota2023erasing} presents a fine-tuning method for concept removal from diffusion models. Additionally, \cite{zhang2023forgetmenot} presents the "Forget-Me-Not" method, which enables the targeted removal of specific objects and content from large models within 30 seconds while minimizing the impact on other content. (2) Dataset Deduplication: \cite{somepalli2022diffusion} explores whether diffusion models create unique artworks or directly replicate certain content from the training dataset during image generation. (3) Machine Unlearning: Numerous model unlearning methods have been developed in the context of image-related tasks \cite{bourtoule2020machine, NEURIPS2019_cb79f8fa, nguyen2022survey}, among others. While machine unlearning is designed to protect the privacy of target samples, \cite{Chen_2021} demonstrates that in the context of model classification tasks, machine unlearning might leave traces.

\textbf{Reinforcement Learning from Human Feedback:} Numerous studies have explored using human feedback to optimize models in various settings, such as simulated robotic control \cite{NIPS2017_d5e2c0ad}, game-playing \cite{2008IEEE}, machine translation \cite{nguyen-etal-2017-reinforcement}, citation retrieval \cite{menick2022teaching}, browsing-based question-answering \cite{song2021denoising}, summarization \cite{stiennon2020learning}, instruction-following \cite{ouyang2022training}, and alignment with specifications \cite{bai2022training}. Recently, \cite{lee2023aligning} studied the alignment of text-to-image diffusion models to human preferences using a method based on reward-weighted likelihood maximization. Their method corresponds to one iteration of the reward-weighted regression (RWR) method. Additionally, \cite{black2023ddpo} proposed a class of policy gradient algorithms to perform reinforcement learning by posing denoising diffusion as a multi-step decision-making problem. Their findings show that DDPO significantly outperforms multiple iterations of weighted likelihood maximization (RWR-style) optimization. Therefore, we adopt the DDPO method for reinforcement learning in this context.

\section{Problem Formulation}

We consider a training dataset $ \mathcal{D} $, composed of copyrighted ($ \mathcal{D}_c $) and non-copyrighted ($ \mathcal{D}_{nc} $) images. The dataset proportions are represented by $ p_{\text{c}} $ and $ p_{nc} $, with $p_{c} + p_{nc} = 1 $.

Each image $ x_i $ in the dataset has an associated feature vector $ \mathbf{f}_i $ and a text prompt $ \mathbf{t}_i $. The model generates an image $ \hat{x}_i $ from $ \mathbf{t}_i $, aiming for $ \hat{\mathbf{f}}_i $ to closely match $ \mathbf{f}_i $. 

To prevent copyright violations, we define a \textit{copyright loss (CL)} that penalizes generated images $ \hat{x}_i $ that are overly similar to those in $ \mathcal{D}_{\text{c}} $. Additionally, the \textit{Fréchet Inception Distance (FID)} evaluates the quality of generated images to ensure they are visually coherent and diverse.

The goal is to train the model to minimize the \textit{copyright loss (CL)}  while maintaining a high FID score, effectively balancing the trade-off between reducing copyright infringement and preserving generative image quality.

\section{Main Approach}

\subsection{Copyright Metric}
We begin by reviewing key aspects of copyright law and introducing the extrinsic and intrinsic legal tests. We then demonstrate how these tests can be analytically modeled using indicators of semantic and perceptual similarity. Finally, we provide a detailed explanation of the methods used to measure semantic and perceptual similarity.

\textbf{Copyright Law.}
In the U.S., proving copyright infringement with AI-generated outputs requires two criteria: the AI must have accessed the copyrighted works, and the outputs must be substantially similar to those works. Courts typically use a two-part test to evaluate substantial similarity: an extrinsic test, which objectively compares specific expressive elements, and an intrinsic test, which subjectively compares the overall impression based on the perception of an ordinary audience. The plaintiff must prove substantial similarity under both tests. Therefore, we aim to select metrics that allow to closely mimic such legal tests.

\textbf{Semantic metric.}
We leverage the CLIP model\cite{radford2021learning} to generate semantic embeddings for anchor and generated images and calculate the metrics by:
\begin{equation} 
\left\{
\begin{aligned}
x &= CLIP(Image_{ori}) \\
y &= CLIP(Image_{gen})
\end{aligned}
\right.
\end{equation}
\begin{equation}
\label{eq:sem}
d_{sem}(x,y) = MSE(x, y),
\end{equation}
where $Image_{ori}$ and $Image_{gen}$ denote the anchor image and generated image, respectively; CLIP denotes the CLIP’s image encoder. Here we utilize the Mean Squared Error (MSE) between the embeddings as the evaluation metric.

\textbf{Perceptual metric.}
Here we used the Learned Perceptual Image Patch Similarity (LPIPS) metric proposed by \cite{lpips}. First, LPIPS normalizes the feature dimension of all pixels and layers to unit length, scales each feature by a specific weight, and we then calculated squared $l_2$ distance between these weighted activations. The squared distances are then averaged across the image dimensions (spatial averaging) and summed over the layers, resulting in the final perceptual distance metric $d$ as follows:
\begin{equation} \label{eq:perc}
d_{perc}(x, y) = \sum_l \frac{1}{H_l W_l} \sum_{i,j} \left\|w^l \odot (\hat{x}^l_{ij} - \hat{y}^{l}_{ij})\right\|_2^2
\end{equation}
where $\hat{x}^l_{ij}$ and $\hat{y}^{l}_{ij}$ denotes the normalized feature vectors at layer $l$ at pixel $(i,j)$. $H_l$ and $W_l$ denote the height and width of the feature map at layer $l$, respectively. The parameter $w^l$ represents the weight assigned to each feature at layer $l$.

We normalize the result and subtract it to 1 to obtain a metric for perceptual similarity $loss_{perc}$ and semantic similarity as follows:
\begin{equation}
CL_{sem}(x,y) = 1 - \frac{d_{sem}(x,y)}{1 + d_{sem}(x,y)}
\end{equation}

\begin{equation}
CL_{perc}(x,y) = \frac{1 - \text{d}_{perc}(x, y)}{1 + \text{d}_{perc}(x, y)}
\end{equation}

Finally, we denote the total copyright loss between a query image y and value image x as:
\begin{equation}
\label{reward}
CL(x,y) = \alpha \cdot CL_{sem}(x,y) + \beta \cdot CL_{perc}(x,y)
\end{equation}

The parameters $\alpha$ and $\beta$ control the trade-off between these two components.

\begin{figure*}[t]
    \centering
    \begin{subfigure}[b]{0.3\textwidth}
        \centering
        \includegraphics[width=\textwidth]{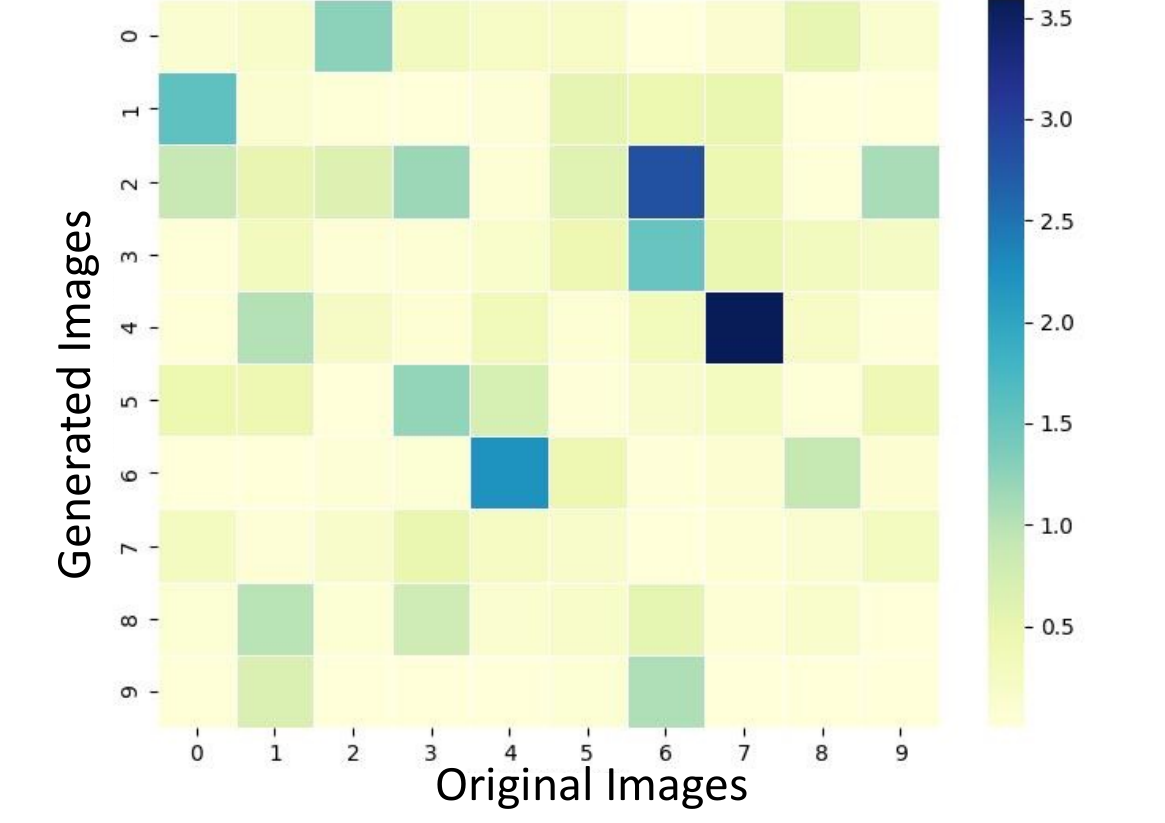}
        \caption{Heatmap of Semantic Scores.}
        \label{fig:sub1}
    \end{subfigure}
    \hfill
    \begin{subfigure}[b]{0.3\textwidth}
        \centering
        \includegraphics[width=\textwidth]{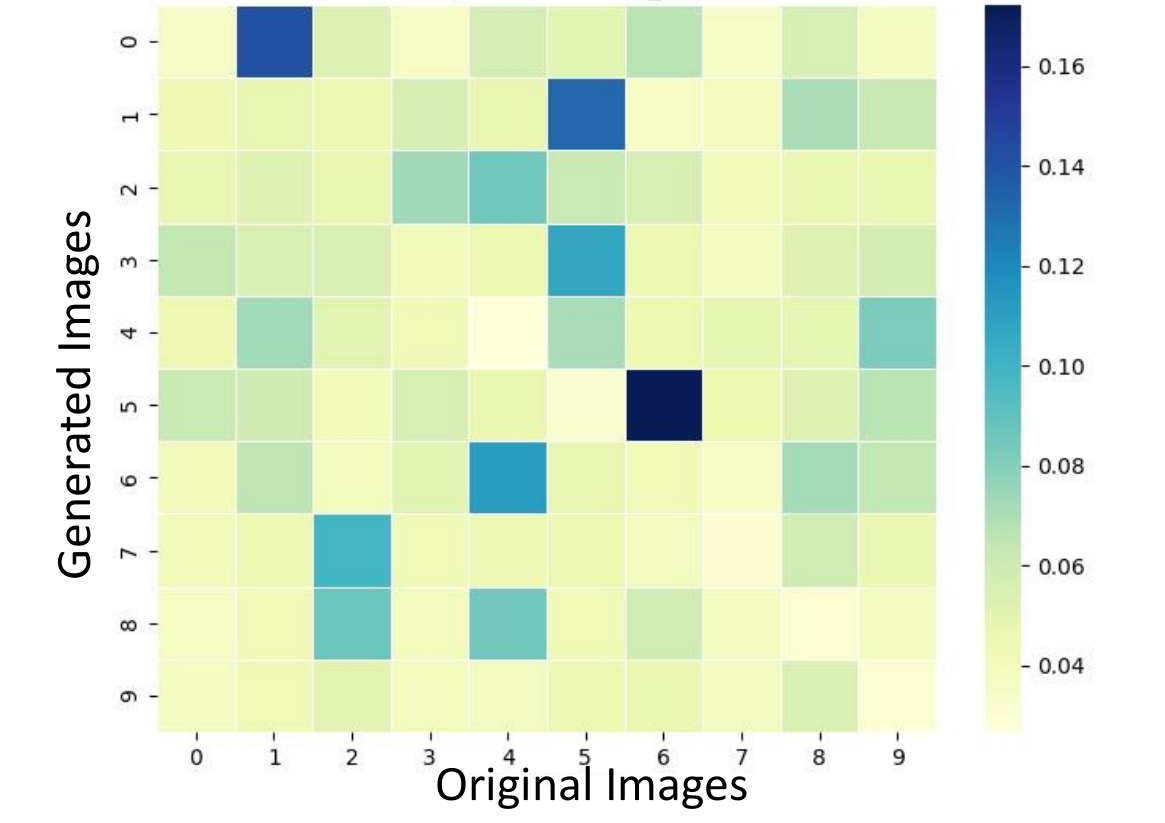}
        \caption{Heatmap of LPIPS Scores.}
        \label{fig:sub2}
    \end{subfigure}
    \hfill
    \begin{subfigure}[b]{0.3\textwidth}
        \centering
        \includegraphics[width=\textwidth]{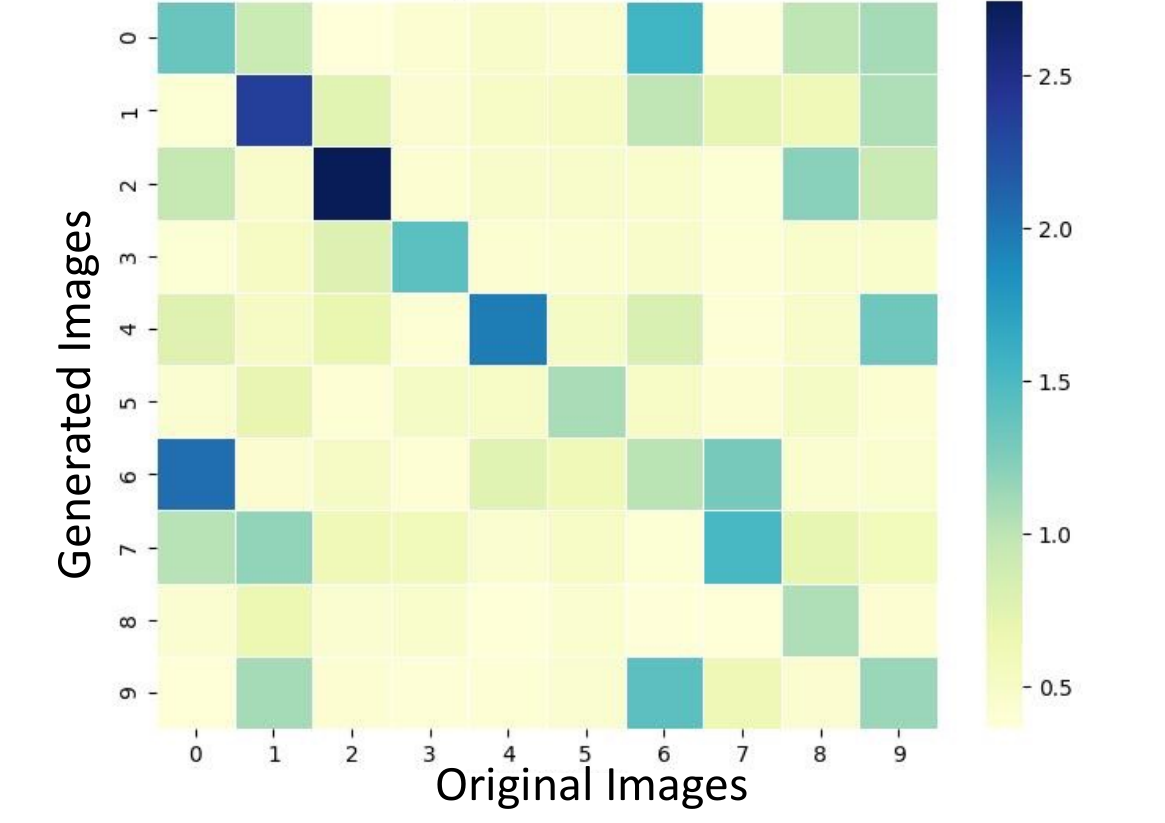}
        \caption{Heatmap of Our CL Scores.}
        \label{fig:sub3}
    \end{subfigure}
     \vspace{-0.1in}
    \caption{Comparison of Similarity Metrics for Detecting Potential copyright infringement.}
    \label{fig:heatmaps}
\end{figure*}

\subsection{The proposed method \methodname{}}

\methodname{} aims to achieve the dual objectives of maintaining high-quality image generation and ensuring compliance with copyright laws, thereby addressing the pressing challenge of copyright infringement in text-to-image diffusion models. The overview progress shows in \ref{fig:overview}, it consists of several key steps:

\subsubsection{Model Initialization}
We start with a pre-trained text-to-image diffusion model $\mathcal{M}_\theta$, where $\theta$ represents the parameters trained on a dataset containing both copyrighted ($\mathcal{D}_{\text{c}}$) and non-copyrighted ($\mathcal{D}_{\text{nc}}$) data. Diffusion models like DDPM approximate the data distribution $q_0(x_0)$ through a reverse Markov chain modeled as:
\begin{equation}
p_\theta(x_{t-1}|x_t) = \mathcal{N}(\mu_\theta(x_t, t), \Sigma_t),
\end{equation}
where the forward diffusion process adds Gaussian noise, and the reverse process iteratively denoises $x_t$ to recover $x_0$.

\subsubsection{Reward Function Design}
We calculate the semantic and perceptual distances $d_{sem}(x_0, c)$ and $d_{perc}(x_0, c)$, using Equations \eqref{eq:sem} and \eqref{eq:perc}, respectively. These metrics are then combined to compute the reward function as follows:
\begin{equation}
r(x_0, c) = \alpha \cdot d_{sem}(x_0, c) + \beta \cdot d_{perc}(x_0, c),
\end{equation}
where a higher similarity between two images results in a lower reward score.
% To guide the model toward copyright-safe generation, we introduce a reward function $r(x_0, c)$ defined as:
% \begin{equation}
% r(x_0, c) = \alpha \cdot d_{sem}(x_0, c) + \beta \cdot d_{perc}(x_0, c),
% \end{equation}
% where:
% \begin{itemize}
%     \item $d_{sem}(x_0, c)$: Semantic similarity loss(e.g., MSE between image embeddings).
%     \item $d_{perc}(x_0, c)$: Perceptual similarity loss (e.g., LPIPS metric).
% \end{itemize}
% The parameters $\alpha$ and $\beta$ balance sensitivity to semantic and perceptual differences, ensuring that generated images deviate sufficiently from copyrighted content.

\subsubsection{Training with DDPO}
The training process leverages Denoising Diffusion Policy Optimization (DDPO), framing the reverse diffusion process as a Markov Decision Process (MDP). The model aims to maximize the expected reward:
\begin{equation}
J(\theta) = \mathbb{E}_{c \sim p(c), x_0 \sim \mathcal{M}_\theta(x_0 | c)} [r(x_0, c)].
\end{equation}
Policy gradient methods iteratively adjust $\theta$ to increase $J(\theta)$, reducing the likelihood of generating infringing content while preserving image quality.

\subsubsection{KL Regularization}
To prevent overfitting to the reward function, KL divergence regularization is added:
\begin{equation}
\mathcal{L}_{\text{KL}} = \sum_{t=1}^{T} \mathbb{E}_{p_\theta(x_t|c)} \left[KL(p_\theta(x_{t-1}|x_t, c) \| p_{\text{pre}}(x_{t-1}|x_t, c))\right],
\end{equation}
penalizing deviations from the original model distribution. The final loss function combines reward optimization with KL regularization:
\begin{equation}
\mathcal{L}(\theta) = -J(\theta) + \lambda \cdot \mathcal{L}_{\text{KL}},
\end{equation}
where $\lambda$ controls the trade-off between compliance and generative fidelity.

\section{Experiment}

In this section, we first evaluate the effectiveness of our proposed copyright loss (CL) metric. Then we evaluate \methodname{} on 3 real-world datasets. Furthermore, we explore the impact of the proportion of copyright images in the training set on the efficiency of \methodname{}.

\subsection{Experiment Setup}

\textbf{Dataset. } Our datasets contain copyright and non-copyright data. To enhance realism, our search is confined to famous artwork and creation figures, which we designate as our copyright dataset:

\begin{enumerate}
    \item \textbf{Paintings: } Painting artworks\cite{vangogh1932} often embody the distinctive style of the artist, encompassing aspects such as brushstrokes, lines, colors, and compositions. We gathered over 1000 paintings from Vincent Van Gogh.
    \item \textbf{Cartoons: } Cartoon figure images, including characters from animations and cartoons, are often protected by law. We have curated a dataset of around 1000 influential animated characters and figures by collecting information from reputable sources such as Kaggle\footnote{\url{https://www.kaggle.com/datasets}} and Wikipedia\footnote{\url{https://www.tensorflow.org/datasets/catalog/wikipedia}}.
    \item \textbf{Portraits: } The right of a portrait encompasses an individual’s authority over their own image, including their facial features, likeness, and posture. We gathered over 500 portrait images from Wikipedia.
\end{enumerate}

For the non-copyright dataset, we sourced images from ImageNet \cite{deng2009imagenet}, selecting one image from each class, resulting in a total of 1,000 images.

\begin{table*}[!t]
\centering
\caption{Comparison of Metrics for Different Datasets.}
\vspace{-0.1in}
\resizebox{0.9\textwidth}{!}{
\begin{tabular}{|l|cccc|cccc|cccc|}
\hline
\multirow{2}{*}{Method} & \multicolumn{4}{c|}{Paintings} & \multicolumn{4}{c|}{Cartoons} & \multicolumn{4}{c|}{Portraits} \\ \cline{2-13} 
                        & CLIP (\%) & CL (\%) & $l_2$-norm (\%) & FID & CLIP (\%) & CL (\%) & $l_2$-norm (\%) & FID & CLIP (\%) & CL (\%) & $l_2$-norm (\%) & FID \\ \hline
SDXL                      & 65.78     & 55.24   & 62.10   & \textbf{13.4}    & 75.56     & 70.23   & 72.21 & \textbf{15.1}       & 60.12     & 52.89   & 58.31 & \textbf{12.7}     \\ 
\methodname{}           & \textbf{35.12} & \textbf{28.45} & \textbf{40.12} & 18.1 & \textbf{30.34} & \textbf{20.76} & \textbf{33.21} & 20.8 & \textbf{30.23} & \textbf{25.67} & \textbf{38.34} & 17.5\\ 
CA                      & 50.34     & 45.23   &52.40  & 14.9     & 65.45     & 60.89   & 62.23   &16.7    & 48.34     & 42.23   & 46.12  & 14.3     \\ 
UCE                     & 52.11     & 48.12   & 56.01  & 14.6     & 67.89     & 63.12   & 65.46  &16.2     & 50.89     & 44.78   & 49.34   &14.1    \\ 
Forget-Me-Not           & 48.76     & 40.67   & 47.42 & 15.1      & 60.33     & 55.47   & 57.34   & 17.0    & 45.56     & 39.12   & 43.27    & 14.8   \\ \hline
\end{tabular}
}
\vspace{-0.1in}
\label{table:comparisonall}
\end{table*}

\begin{figure*}[t]
\centering
\begin{subfigure}[b]{0.3\textwidth}
    \centering
    \includegraphics[width=\textwidth]{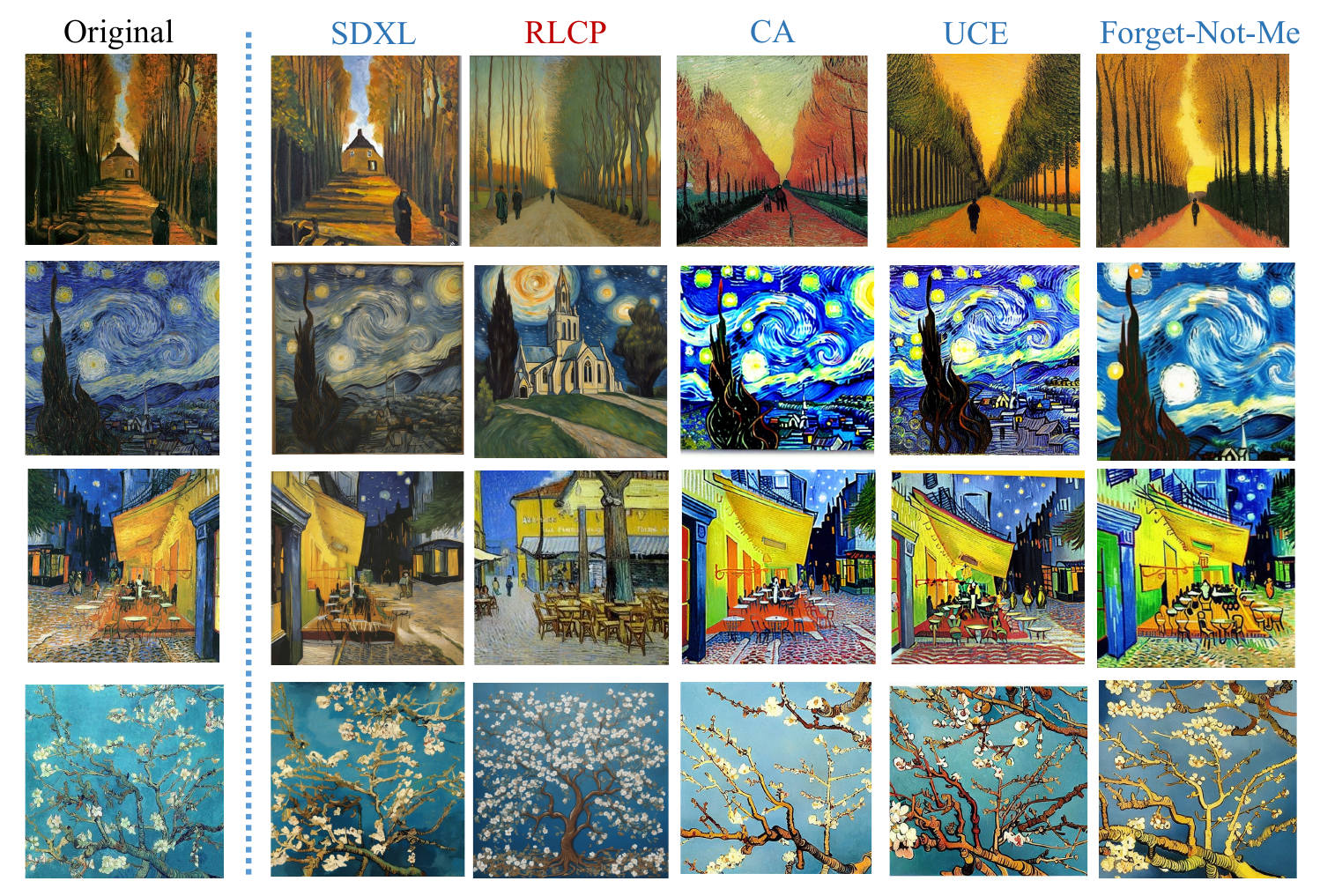}
    \caption{Paintings}
    \label{fig:paintings_example}
\end{subfigure}
\hfill
\begin{subfigure}[b]{0.3\textwidth}
    \centering
    \includegraphics[width=\textwidth]{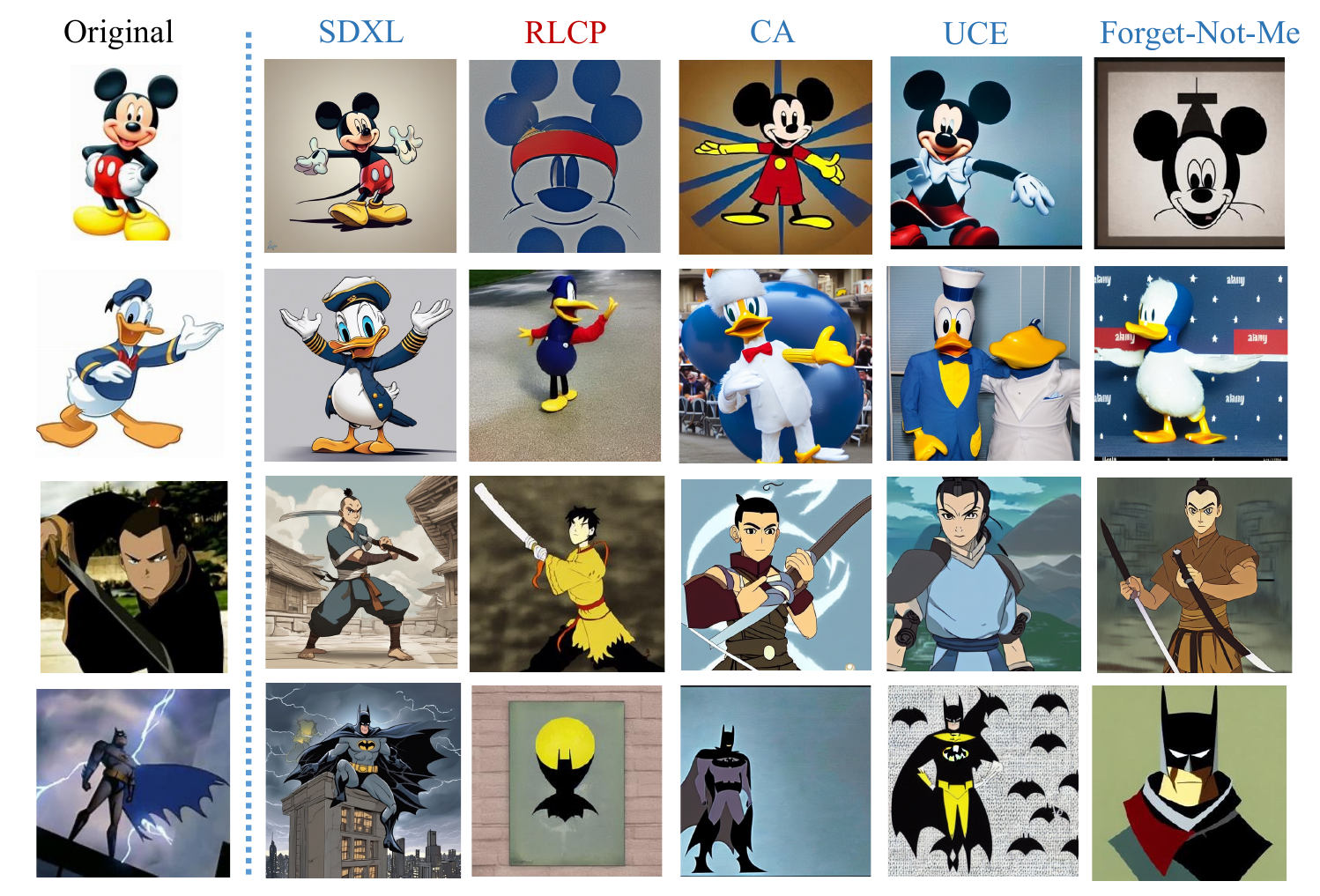}
    \caption{Cartoons}
    \label{fig:cartoon_example}
\end{subfigure}
\hfill
\begin{subfigure}[b]{0.3\textwidth}
    \centering
    \includegraphics[width=\textwidth]{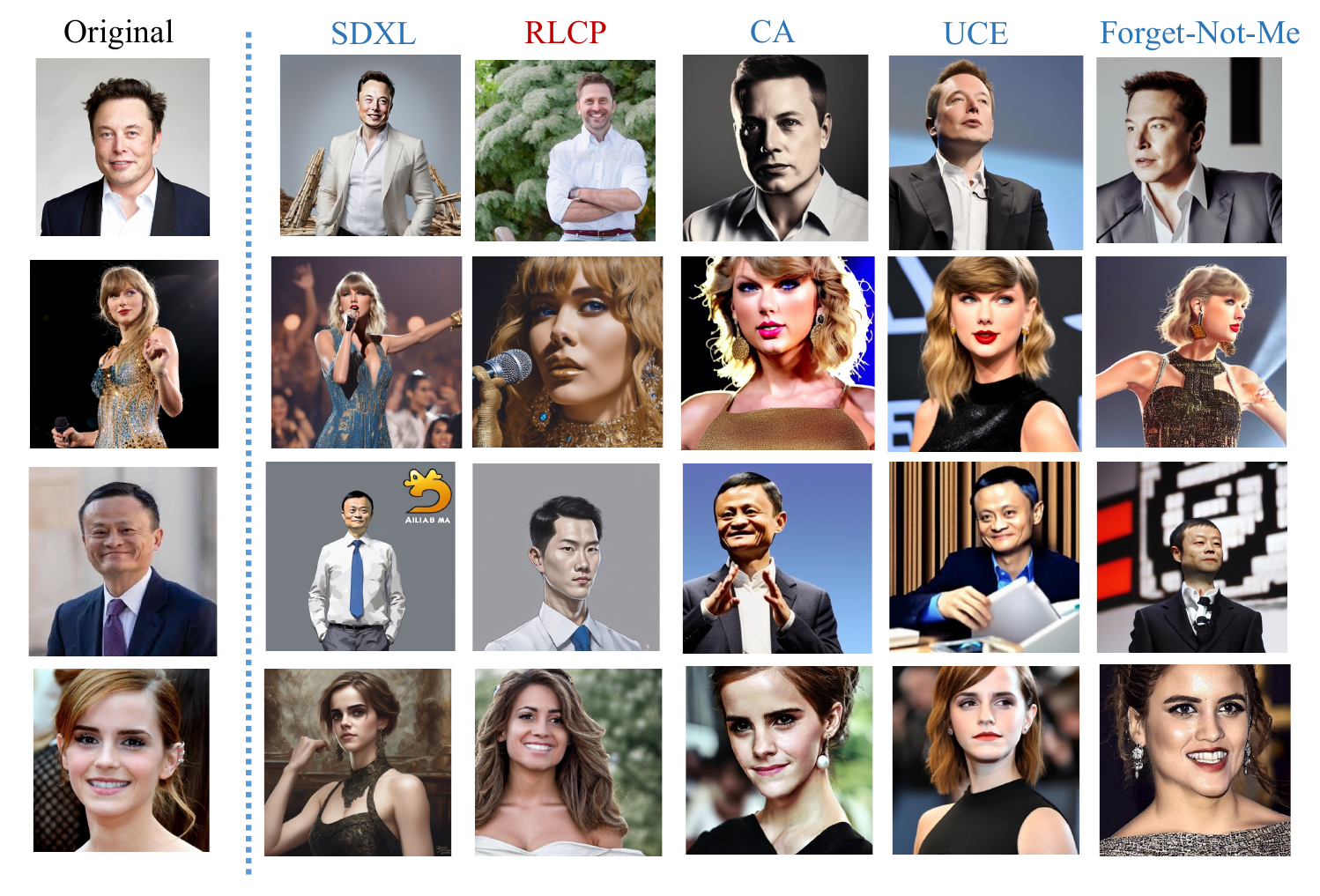}
    \caption{Portraits}
    \label{fig:portrait_example}
\end{subfigure}
\vspace{-0.1in}
\caption{Examples of Fine-tuned Results for Copyright Protection Across Different Datasets.}
\label{fig:all_examples}
\end{figure*}

\textbf{Evaluation Metric. } We use our proposed metric copyright loss (CL) as well as CLIP and $l_2$-norm to measure the degree of copyright violation. We also use FID to measure the generative quality of text-to-image diffusion model. 
\begin{enumerate}
    \item \textbf{CLIP: } We evaluate CLIP score changes for text-image similarity.
    \item \textbf{CL: } We quantify the similarity between the original copyright images and their unlearned counterparts utilizing our proposed copyright loss (CL) metric.
    \item \textbf{$l_2$-norm: } We calculate the squared $l_2$ distance between generated and training images. 
    \item \textbf{Fréchet Inception Distance (FID): } We use FID to evaluate the generative image quality of text-to-image diffusion model.
%     The formulation of the metric is as follows:
% \begin{equation}
% \text{FID} = \|\mu_x - \mu_y\|^2 + \text{Tr}(\Sigma_x + \Sigma_y - 2(\Sigma_x \Sigma_y)^{1/2})
% \end{equation}
% where $\mu_x$ and $\Sigma_x$ are the mean and covariance matrix of the feature vectors from real images, $\mu_y$ and $\Sigma_y$ are the mean and covariance matrix of the feature vectors from generated images, and $\text{Tr}(\cdot)$ denotes the trace of a matrix.
% where:
% \begin{itemize}
%     \item \(\mu_x\) and \(\Sigma_x\) represent the mean and covariance matrix of the feature vectors from real images.
%     \item \(\mu_y\) and \(\Sigma_y\) represent the mean and covariance matrix of the feature vectors from generated images.
%     \item \(\text{Tr}(\cdot)\) denotes the trace of a matrix.
% \end{itemize}
\end{enumerate}

This evaluation provides valuable insights into assessing copyright infringement while preserving the ability to generate non-infringing content.

\textbf{Baselines. } Four baselines are listed as follows:
\begin{itemize}
    \item \textbf{SDXL: } SDXL\cite{von-platen-etal-2022-diffusers} is a latent text-to-image diffusion model capable of generating photo-realistic images given any text input.
    \item \textbf{Concept Removal (CA): }  CA\cite{gandikota2023erasing} An efficient method of ablating concepts in the pretrained model.
    \item \textbf{Unified Concept Editing (UCE): } UCE\cite{kumari2023ablatingconceptstexttoimagediffusion} edits the model without training using a closed-form solution, and scales seamlessly to concurrent edits on text-conditional diffusion models.
    \item \textbf{Forget-Me-Not: } Forget-Me-Not\cite{zhang2023forgetmenot} is an efficient solution designed to safely remove specified IDs, objects, or styles from a well-configured text-to-image model.
    % \item k-NAF: A generative model learning algorithm, which efficiently modify the original generative model learning algorithm in a black box manner, that outputs generative models with strong bounds on the probability of sampling protected content.
\end{itemize}

% \begin{table}[th]
%     \centering
%     \caption{Hyperparameters configuration}
%     \begin{tabular}{|l|l|} \hline
%         batch\_size & 8\\
%         learning\_rate & 3$e^{-4}$\\
%         Samples per iteration & 32\\
%         Gradient updates per iteration & 4\\
%         Clip range & $e^{-4}$\\ \hline
%     \end{tabular}
%     \vspace{-0.1in}
    
%     \label{tab:hyperparams}
% \end{table}

% \begin{figure}[t]
% \centering
% \includegraphics[width=0.9\columnwidth]{images/painting.pdf}
%  \vspace{-0.1in}
% \caption{Examples of fine-tuned results for copyright protection on Paintings.}
% \label{fig:paintings_example}
% \end{figure}

% \begin{figure}[t]
% \centering
% \includegraphics[width=0.9\columnwidth]{images/cartoon.pdf}
%  \vspace{-0.1in}
% \caption{Examples of fine-tuned results for copyright protection on Cartoon images.}
% \label{fig:cartoon_example}
% \end{figure}

% \begin{figure}[t]
% \centering
% \includegraphics[width=0.9\columnwidth]{images/portrait.pdf}
%  \vspace{-0.1in}
% \caption{Examples of fine-tuned results for copyright protection on Portrait images.}
% \label{fig:Portrait_example}
% \end{figure}

\textbf{Experimental environment and hyperparameters. } All the experiments were conducted on a cluster with 2 80Gb A100 GPUs. The hyperparameters used were: a batch size of 8, a learning rate of $3 \times 10^{-4}$, 32 samples per iteration, 4 gradient updates per iteration, and a clip range of $10^{-4}$.

\begin{figure*}[!t]
    \vspace{-0.1in}
    \centering
    \begin{subfigure}[b]{0.32\textwidth}
        \centering
        \includegraphics[width=\textwidth]{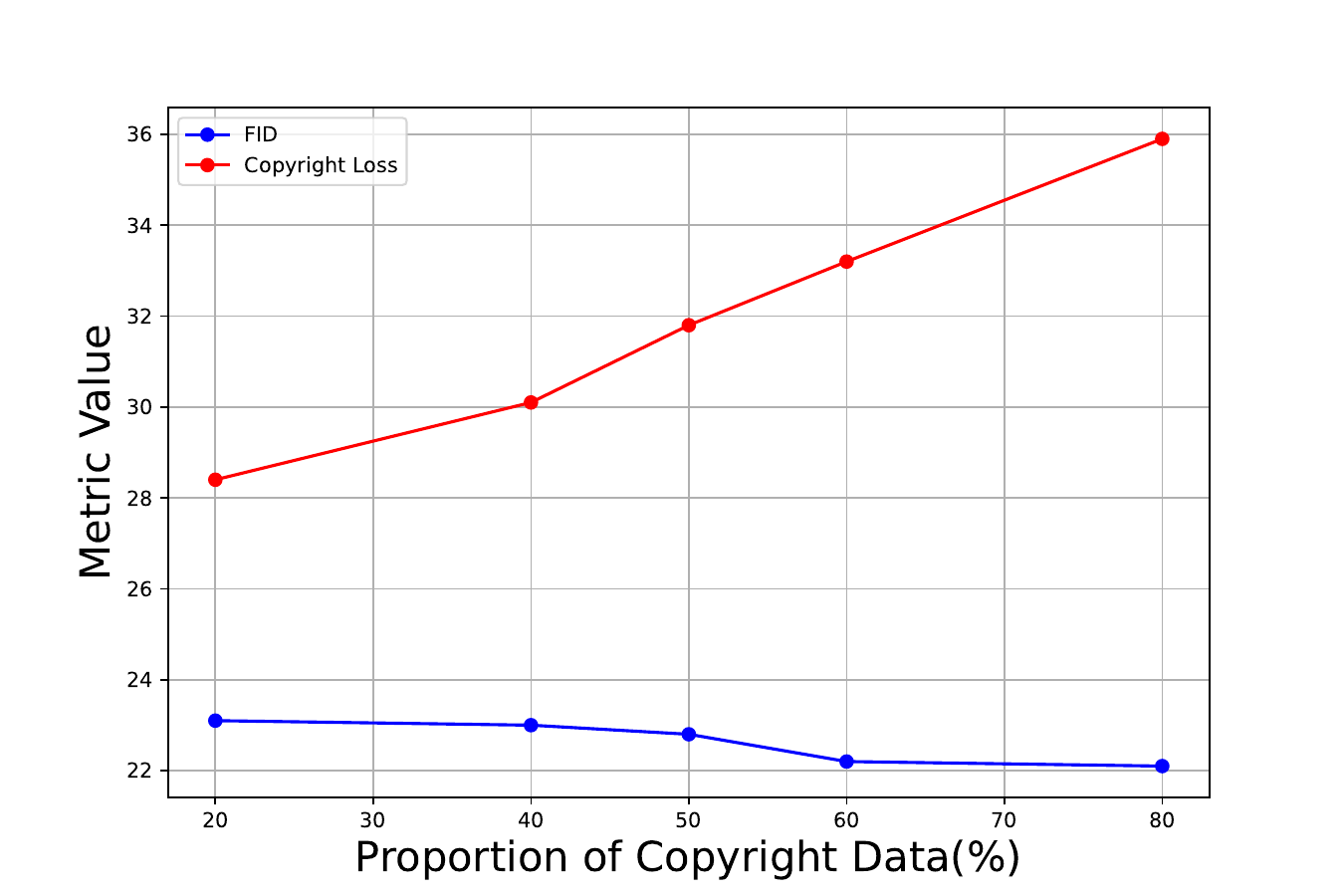}
        \caption{Paintings.}
        \label{fig:painting_pro}
    \end{subfigure}
    \hfill
    \begin{subfigure}[b]{0.32\textwidth}
        \centering
        \includegraphics[width=\textwidth]{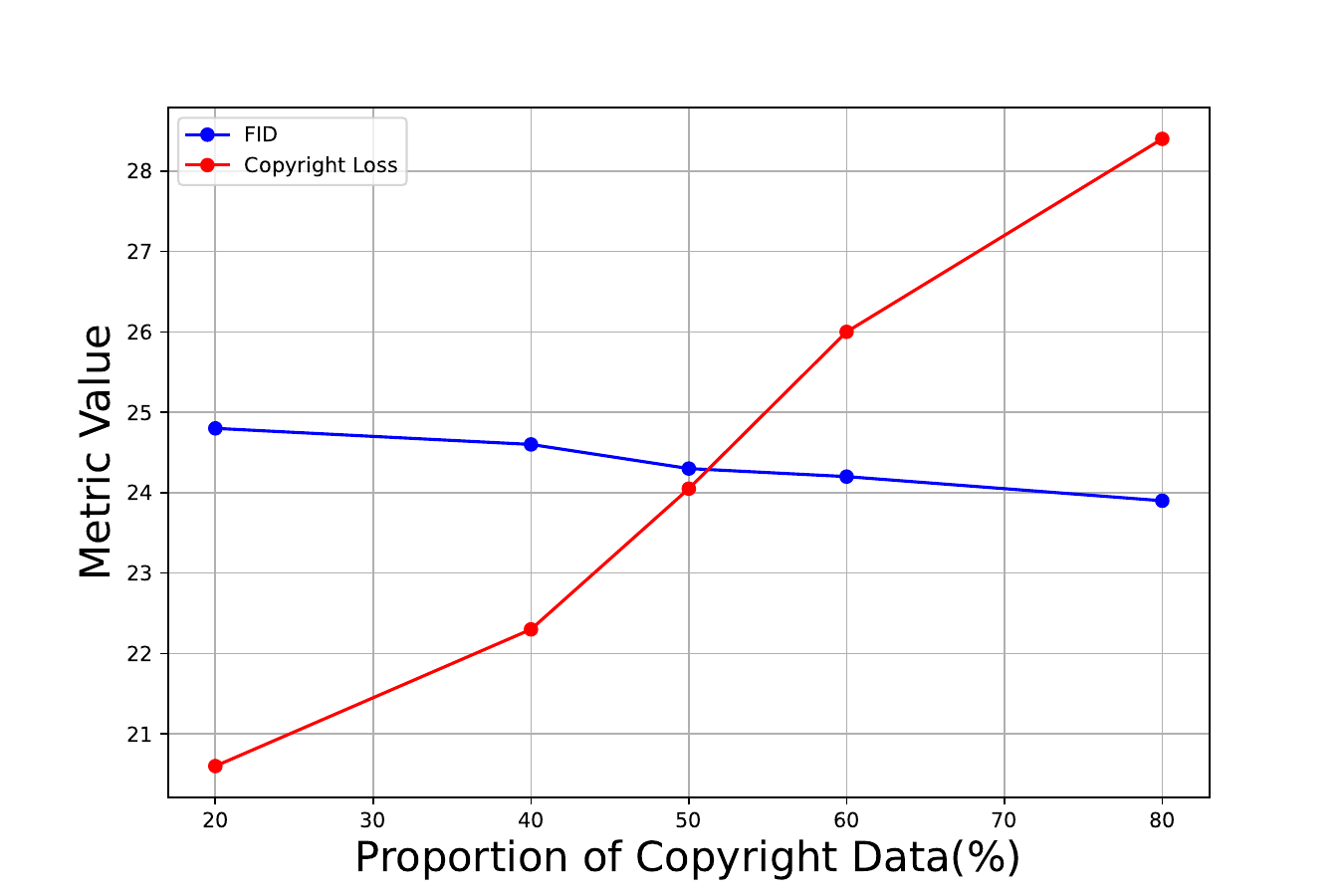}
        \caption{Cartoons.}
        \label{fig:cartoon_pro}
    \end{subfigure}
    \hfill
    \begin{subfigure}[b]{0.32\textwidth}
        \centering
        \includegraphics[width=\textwidth]{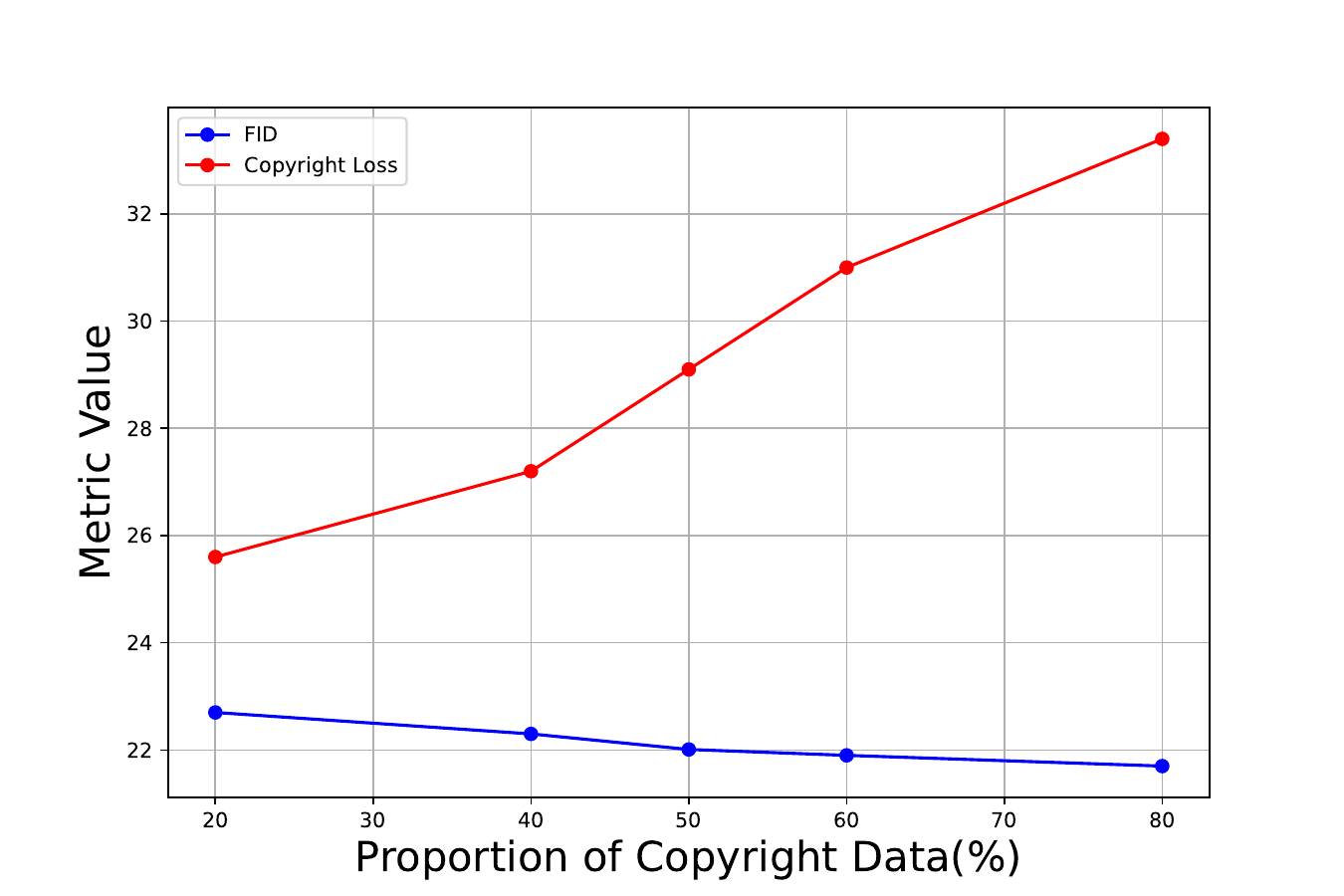}
        \caption{Portraits.}
        \label{fig:portrait_pro}
    \end{subfigure}
    \vspace{-0.1in}
    \caption{Impact of Copyright Data Proportion on Copyright Loss and FID for Different Datasets.}
    \label{fig:proportion}
\end{figure*}

\subsection{Experiment Results}
We evaluate the effectiveness of our proposed \methodname{} method across several dimensions. We compare our results against baselines, analysis of individual metrics, and analyze the trade-offs between copyright protection and image quality.

\textbf{Effectiveness of copyright metric. }
To validate the reliability of our copyright loss (CL) metric, we conducted experiments on 10 copyrighted paintings. CLIP embeddings were used to compute semantic similarity via Mean Squared Error, and perceptual similarity was measured using LPIPS. 

In Figure \ref{fig:heatmaps}, each heatmap reflects the similarity scores between each pair of images, with the diagonal cells highlighting scores between identical pairs. The “query index” refers to the position of the query image in the dataset, while the “value index” represents the corresponding value or similarity score for each generated image. The first two heatmaps show the similarity scores using individual metrics, where image similarity is represented, but the clarity and emphasis on identical pairs are less pronounced. In the third heatmap, which uses the combined metric, the similarity is more distinctly captured, especially along the diagonal, where the color is noticeably deeper and more defined. This shows that the combined metric provides a clearer and more accurate representation of the similarity between the images.

\begin{table*}[!t]
\centering
\caption{Ablation Study Results}
\vspace{-0.1in}
\resizebox{0.9\textwidth}{!}{
\begin{tabular}{|l|cccc|cccc|cccc|}
\hline
\multirow{2}{*}{Method} & \multicolumn{4}{c|}{Paintings}          & \multicolumn{4}{c|}{Cartoons}          & \multicolumn{4}{c|}{Portraits}          \\ \cline{2-13} 
                        & CLIP (\%) & CL (\%) & $l_2$-norm (\%) & FID   & CLIP (\%) & CL (\%) & $l_2$-norm (\%)  & FID     & CLIP (\%) & CL (\%) & $l_2$-norm (\%) & FID      \\ \hline
\methodname{}$_{semantic}$       & 47.12     & 43.78   & 55.12     & 15.0       & 47.45     & 38.12   & 47.54     & 17.6  & 47.34     & 44.78   & 53.12     & 16.8         \\ 
\methodname{}$_{perceptual}$     & 43.45     & 39.12   & 51.12     & 17.5      & 43.56     & 30.12   & 40.78     & 19.8  & 41.45     & 38.78   & 49.56     & 17.2       \\ \cline{1-13}  
\methodname{}                    & \textbf{35.12}     & 28.45   & 40.12     & 18.1      & 30.34     & 20.76   & 33.21     & 20.8  & 30.23     & 25.67   & 38.34    & 17.5       \\ \hline
\end{tabular}
}
\label{table:ablation}
\end{table*}

\textbf{Performance of Copyright Protection. }
% Our proposed method, \methodname{}, is compared against the baselines to evaluate its ability to minimize copyright infringement while maintaining the quality of generated images. The results are listed in Table \ref{table:comparisonall}. 
The comparison of \methodname{} with baselines is shown in Table \ref{table:comparisonall}. While \methodname{} shows relatively weaker performance on the FID, it still performs exceptionally well on the $l2$-norm metric. This discrepancy can be explained by the nature of FID, which measures the distribution similarity between generated and target images. Since \methodname{} focuses on avoiding the generation of copyrighted data, it naturally alters the distribution, leading to a higher FID score. Despite this, the $l2$-norm results indicate that \methodname{} maintains strong performance in generating non-infringing content that aligns with the original image features. We also give examples of fine-tuned results for copyright protection on three datasets in Figure \ref{fig:all_examples}.

% \begin{table}[!t]
% \centering
% \caption{Ablation Study.}
% \large
% \resizebox{\columnwidth}{!}{
% \begin{tabular}{|l|c|c|c|c|}
% \hline
% Metric           & SD  & RLCP_{semantic} & RLCP_{perceptual}  & RLCP \\ \hline
% CLIP(\%)             & 75.56   & 45.12              & 38.45                         & 31.23     \\ \hline
% CL(\%)               & 70.23   & 40.78              & 35.12                           & 24.96     \\ \hline
% $l_2$-norm(\%)       & 72.31    & 50.32               & 45.21                            & 37.41      \\ \hline
% FID              & 15.5    & 18.3               & 19.2                         & 18.8      \\ \hline
% \end{tabular}
% }
% \vspace{-0.1in}
% \label{table:ablation}
% \end{table}
% \begin{table}[!t]
% \centering
% \caption{Performance Comparison of using Single Metric}
% \resizebox{0.95\columnwidth}{!}{
% \begin{tabular}{|l|c|c|c|c|c|c|}
% \hline
% Metric           & SD   & Semantic-Only  &  Perceptual-Only  & CA   & UCE  & Forget-Me-Not \\ \hline
% CLIP             & 75.56 & xx & xx & 65.45 & 67.89 & 60.33 \\ \hline
% CL               & 70.23 & xx & xx & 60.89 & 63.12 & 55.47 \\ \hline
% $l_2$-norm          & 0.72 & xx & xx & 0.62 & 0.65 & 0.57 \\ \hline
% FID              & 15.5  & xx & xx  & 16.7  & 16.2  & 17.0  \\ \hline
% \end{tabular}
% }
% \vspace{-0.1in}

% \label{table:ablation}
% \end{table}

\textbf{Ablation Study.} 
To evaluate the impact of individual metrics in the reward function, we assessed the model using only semantic similarity (CLIP) or perceptual similarity (LPIPS) into Reinforcement Learning framework, excluding the combined metric. The results are shown in Table \ref{table:ablation}.

The $\methodname{}_{semantic}$ approach achieves higher CLIP scores compared to $\methodname{}_{perceptual}$, indicating better semantic alignment with the target content. However, it underperforms in metrics like $l_2$-norm, where Perceptual-Only outperforms Semantic-Only across all categories. This suggests that Perceptual-Only generates images with features more visually distinct from copyrighted content.

The combined metric (\methodname{}) balances these trade-offs, achieving the lowest CLIP and $l_2$-norm scores, while also demonstrating better FID performance compared to $\methodname{}_{semantic}$ and $\methodname{}_{perceptual}$. Integrating both metrics provides stronger copyright protection and higher generative quality than using either metric individually.
% To assess the impact of individual metrics, we evaluated the model using only semantic (CLIP) or perceptual (LPIPS) similarity.

% The results of the ablation study are presented in Table \ref{table:ablation}. Semantic-Only performs worse than Perceptual-Only, but the combined metric, integrating both semantic and perceptual metrics, providing strong copyright protection while maintaining high generative quality. This confirms the effectiveness of our approach in leveraging both metrics to address the trade-off between copyright compliance and content quality.

\textbf{Impact of Copyright Data Proportion. }
Figure \ref{fig:proportion} illustrates how varying the proportion of copyrighted data impacts \methodname{}’s performance. As the proportion of copyrighted data increases, copyright loss (CL) rises due to the model generating images that more closely resemble the copyrighted content. Conversely, FID decreases slightly, reflecting improved image quality.

These results address concerns about scalability, showing that \methodname{} performs robustly even with a high concentration of copyrighted data. A lower FID score indicates better image quality, but this comes at the cost of increased copyright loss (CL). This trade-off is particularly evident when training data contains highly similar copyrighted images.

\section{Conclusion}
In this paper, we presented a Reinforcement Learning-based Copyright Protection (\methodname{}) for copyright infringement in text-to-image diffusion model. \methodname{} proposes a copyright loss metric that mirrors legal tests used to assess substantial similarity, and then integrates this metric into a reinforcement learning framework for model fine-tuning, and the use of KL divergence to regularize and stabilize the model training process. Experiments conducted on three mixed datasets of copyright and non-copyright images show that \methodname{} significantly reduces the likelihood of generating infringing content while preserving the visual quality of the generated images.

% \begin{table}[htbp]
% \caption{Table Type Styles}
% \begin{center}
% \begin{tabular}{|c|c|c|c|}
% \hline
% \textbf{Table}&\multicolumn{3}{|c|}{\textbf{Table Column Head}} \\
% \cline{2-4} 
% \textbf{Head} & \textbf{\textit{Table column subhead}}& \textbf{\textit{Subhead}}& \textbf{\textit{Subhead}} \\
% \hline
% copy& More table copy$^{\mathrm{a}}$& &  \\
% \hline
% \multicolumn{4}{l}{$^{\mathrm{a}}$Sample of a Table footnote.}
% \end{tabular}
% \label{tab1}
% \end{center}
% \end{table}

% \begin{figure}[htbp]
% \centerline{\includegraphics{fig1.png}}
% \caption{Example of a figure caption.}
% \label{fig}
% \end{figure}

\vspace{-0.1in}
\bibliographystyle{IEEEbib}
\bibliography{icme2025}

\end{document}